# Comparison of Different Segmentations in Automated Detection of Hypertension Using Electrocardiography with Empirical Mode Decomposition

Y.E. ERDOĞAN [1, 2], A. NARİN [2] and W. HARIRI [3]

[1] Eregli Iron and Steel Co., Zonguldak, Turkey,
yeerdogan@erdemir.com.tr
[2] Electrical and Electronics Engineering Department, Zonguldak Bülent Ecevit University, 67100, Zonguldak, Turkey, alinarin45@gmail.com
[3] Labged Laboratory, Computer Science department, Badji Mokhtar Annaba University, Annaba, Algeria, hariri.walid@hotmail.com

*Abstract* Hypertension (HPT) refers to a condition where the pressure exerted on the walls of arteries by blood pumped from the heart to the body reaches levels that can lead to various ailments. Annually, a significant number of lives are lost globally due to diseases linked to HPT. Therefore, the early and accurate diagnosis of HPT is of utmost importance. This study aimed to automatically and with minimal error detect patients suffering from HPT by utilizing electrocardiogram (ECG) signals. The research involved the collection of ECG signals from two distinct groups. These groups consisted of ECG data of both five thousand and ten thousand data points in length, respectively. The performance in HPT detection was evaluated using entropy measurements derived from the 5-layer Intrinsic Mode Function (IMF) signals through the application of the Empirical Mode Decomposition method. The resulting performances were compared based on the nine features extracted from each IMF. To summarize, employing the 5-fold cross-validation technique, the most exceptional accuracy rates achieved were 99.9991% and 99.9989% for ECG data of lengths five thousand and ten thousand, respectively, using decision tree algorithms. These remarkable performance results indicate the potential usefulness of this method in assisting medical professionals to identify individuals with HPT.

## I. Introduction

Hypertension (high blood pressure) is a health concern characterized by excessive pressure within the arteries [1]. In our country, hypertension affects one in every three adults [2]. If left untreated, hypertension can lead to various health problems and organ damage. Among the root causes of hypertension, hereditary factors and high salt consumption take the lead in research. However, the precise cause of hypertension in most patients remains unknown [3]. Hypertension is a condition that requires careful management; sudden spikes in blood pressure can result in conditions like stroke and even cerebral hemorrhage [4]. Common symptoms of hypertension include headaches, dizziness, shortness of breath, palpitations, chest pain, and vision issues [5]. One prominent diagnostic method for hypertension is monitoring ambulatory blood pressure over a 24-hour period [6]. However, this method may not effectively detect certain forms of hypertension. Therefore, additional measurements such as electrocardiography (ECG) and echocardiography are necessary for hypertension diagnosis. Because hypertension can significantly affect a person's daily life and quality of life, early and accurate diagnosis is critically important. In the scientific literature, there are studies focused on the automatic detection of hypertension using biomedical signal processing and machine learning methods. Khan and colleagues conducted a study to identify individuals with hypertension using pulse plethysmography signals. They performed feature extraction using Empirical Mode Decomposition (EMD) and demonstrated their detection performance using the k-Nearest Neighbors (k-NN) classifier. Engaging the k-NN algorithm, they achieved a remarkable 99.4% accuracy rate [7]. Rajput and colleagues used ECG signals to identify individuals with hypertension. They extracted features using non-linear methods like wavelet entropy and sample entropy, along with the 5-level wavelet transform. They reported performance results using k-NN, Support Vector Machines (SVM), Ensemble Bagged Tree (EBT), and decision tree algorithms. The highest accuracy, 99.95%, was achieved using the EBT algorithm [8]. In another study, Soh and colleagues focused on the automatic detection of hypertension patients using ECG signals. They used Convolutional Neural Network (CNN) models and achieved a remarkable 99.99% accuracy [9]. Poddar and colleagues used heart rate variability (HRV) signals to identify individuals with hypertension. They worked with time domain, frequency domain, and non-linear entropy measurements. Utilizing the DVM algorithm, they achieved a remarkable 100% accuracy rate [10].

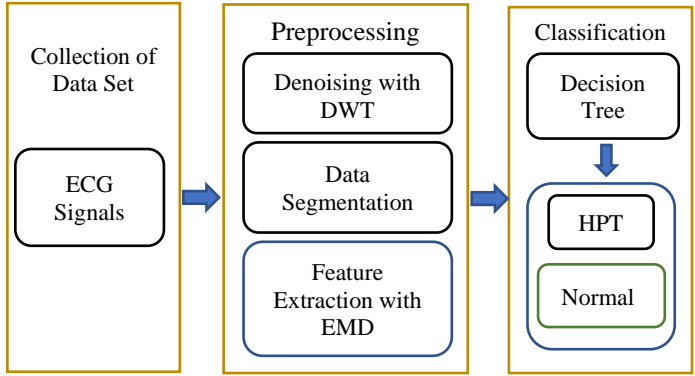

**Figure 1:** Flow Diagram of the Method Proposed in the Study

In this study, publicly available data obtained free of charge from the Physionet.org website was used to make a healthy comparison with studies in the literature. The data set includes ECG data from two classes: healthy and HPT patients. Unlike other studies, ECG signals were examined with two different approaches. These approaches are respectively; They are 5000 and 10000 length ECG signals. 5-layer Internal Mode Functions (IMF) were obtained using the EMD method for each signal type. Time domain characteristics from each IMF signal included the utilization of standard deviation, mean, and root mean square. Non-linear chaotic assessments encompassed the application of measurements such as Shannon entropy, log energy, threshold, time, norm, and approximate entropy. The flow diagram is given in detail in Figure 1. In the following sections of the study, the adjustment of the data used, the pre-processes applied on the data, the measurements used for feature extraction, classifier illumination, performance demonstrations are mentioned, and the final data, the results obtained and the discussion are included.

## II. MATERIAL AND METHODS

### A. Dataset

The data on HPT patients and normal individuals given in Table 1 are available on publicly accessible Phsionet.org. The ECG data of HPT patients is taken from the "Smart Health for Assessing the Risk of Events via ECG (SHAREE) database" [11]. HPT treatment consists of 139 ECG data from people aged 55 and over, 49 of whom are women and 90 are men. Running the ECG belonging to the class that can be kept in a normal environment, the "Massachusetts Institute of Technology-Beth Israel Hospital (MIT-BIH) normal sinus rhythm database" data can be run. It consists of 18 ECG signals in total [12]. A recording consists of approximately 24-hour periods. These recordings were sampled with 128 samples with 8-bit images.

**Table 1.** Data Used in the Study and Their Characteristics

| Class | Number of Data | Female | Male | Age Range | Data Length |
|---|---|---|---|---|---|
| **HPT** | 139 | 49 | 90 | 46-92 | 1.000.000 |
| **Normal** | 18 | 13 | 5 | 20-50 | 1.000.000 |

The first parts of each HPT signal between 0 and 20000 were excluded from the study because they contained a distorted ECG signal.

### B. Noise Elimination with Discrete Wavelet Transform (DWT)

In order to purify the ECG signal from noise, the process of decomposing the noisy signal is started by applying wavelet transform. Thanks to the wavelet transform, it is possible to separate the signal into coefficient groups at different frequency levels. Understanding how the signal behaves in different frequency segments will allow choosing the most appropriate threshold in the next stage. The next step is to determine the best threshold values and eliminate unwanted data by applying threshold values to these coefficients. The last step of the algorithm is to reconstruct the ECG signal using filtered coefficient sets.

### C. Data Segmentation

In this study, each signal consisting of one 980 thousand long data was divided into ten thousand and five thousand length parts and used, respectively. At the end of this process, data for 1800 healthy and 13622 HPT patients were created for ten thousand length data. Also, data for 3600 healthy and 27244 HPT patients were created for five thousand length data.

### D. Empirical Mode Decomposition

EMD stands as a suitable method for examining both stationary and non-linear data sequences, primarily focusing on local-level oscillatory signals [13,14]. Within this algorithm, the Intrinsic Mode Functions (IMFs) serve as the cornerstone. The EMD process commences with the identification of local signal peaks. Once located, these peaks become the foundation for constructing overlapping curves, achieved through 3rd-degree polynomials. Simultaneously, a sub-winding curve is formed based on local signal minima. Subsequently, the mean values of the upper and lower winding curves are computed and subtracted from the original signal, effectively filtering out the low-frequency component. If the resulting signal qualifies as an IMF, the process concludes. However, if it doesn't meet IMF criteria, the algorithm restarts for the new signal. The determination of IMF eligibility hinges on specific conditions. First, the count of zero crossings in the signal should either match or exceed the number of peaks. Secondly, symmetry between the winding curves, established through local minima and maxima, must be satisfied. Time-related attributes extracted from each IMF signal involved the use of standard deviation, mean, and root mean square. The assessment of non-linear, chaotic features involved employing measurements such as Shannon entropy, log energy, threshold, time, norm, and approximate entropy.

### E. Decision Trees

A decision tree is composed of nodes and connections. The tree initiates from a node known as the root node, which serves as the starting point and does not have any incoming connections. Nodes leading to other nodes are referred to as internal nodes or decision nodes, while all other nodes are termed as terminal or leaf nodes. Internal nodes segment input data into two or more subsets based on specific discrete functions of the input features. Furthermore, each leaf node can contain a probability vector that represents the likelihood of the target feature having a particular value. Following the determinations made by the internal nodes, data samples are guided through the tree from the root to a leaf node for classification.

## F. Performance Metrics

In this study, the outcomes were evaluated utilizing five distinct performance measures [15-18]. These:

$$Accuracy(Acc) = \frac{TP + TN}{TP + FN + FP + TN} \quad (1)$$

$$Recall(Rec) = \frac{TP}{TP + FN} \quad (2)$$

$$Specifity(Spe) = \frac{TN}{TN + FP} \quad (3)$$

$$Precision(Pre) = \frac{TP}{TP + FP} \quad (4)$$

$$F1 - score(F1) = \frac{2 * PRE * REC}{PRE + REC} \quad (5)$$

In the context of classification, TP (True Positives) represents the count of individuals correctly identified as having HPT by the classifier. FN (False Negatives) denotes the count of individuals inaccurately classified as normal when they actually have HPT. TN (True Negatives) represents the count of individuals correctly identified as normal by the classifier when they are indeed normal. Lastly, FP (False Positives) signifies the count of individuals mistakenly labeled as having HPT when, in fact, they do not have the condition [19].

## III. EXPERIMENTAL RESULTS

This research study involved the execution of all procedures through MATLAB 2021a software. Subsequently, 45 measures based on Intrinsic Mode Functions (IMF) were derived. The comparison of various segmentations is presented in Table 1. As illustrated in Table 1, the most notable accuracy rate, reaching 99.9991%, was achieved when employing Decision Tree on ECG data with a length of five thousand.

## IV. DISCUSSION

The results were compared with 9 time domain and nonlinear measurements obtained from 5 different IMF signals. Results were obtained using two different segmentations. In order to achieve more reliable and stable results, all data were trained and tested with the 5-fold cross-validation method. The performance values found by using the features taken from all IMF measurements for denoised signals are shown in Table 2. In these studies, traditional machine learning approaches are mostly used. From Table 3. Considering the results obtained, Khan et al. applied the Empirical Mode Decomposition (EMD) technique in conjunction with the k-Nearest Neighbors (k-NN) classifier. Their results demonstrated an accuracy of 99.4% as reported in [7]. Rajput et al. used wavelet transform and they achieved 99.95% accuracy by using the method and k-NN, SVM, EBT classifiers. [8]. Soh et al. achieved 99.99% accuracy using CNN and fully connected layer (FCL) [9]. Poddar et al. used time domain measurements and FFT measurements over HRV signals and 100% accuracy value was achieved using the SVM classifier [10]. In this study, each data was increased by dividing it into ten thousand and five thousand length segments. Data obtained without noise used. Observing Figure 2, it becomes evident that 14 individuals in the HPT category were identified erroneously, whereas 27230 were accurately recognized. In contrast, 13 individuals in the normal category were misclassified, while 3587 were correctly identified. With the features obtained from the 5-level IMFs obtained by the EMD method, an accuracy value of 99.9991% was achieved using decision trees algorithms via five thousand (5000) length signals. Additionally, as depicted in Figure 3, it's evident that 7 individuals belonging to the HPT category were mistakenly identified, while 13615 were accurately recognized. Conversely, 9 individuals from the normal category were misclassified, whereas 1791 were correctly identified. With the features obtained from the 5-level IMFs obtained by the EMD method, an accuracy value of 99.9989% was achieved using decision trees algorithms via ten thousand (10000) length signals. From this point of view, it can be understood that if data segmented smaller parts, it will increase the accuracy. Additionally, working with short data eases the computational burden. It is also thought that it will reduce the stress of patients during data collection. It has been shown that it provides more successful results than many similar studies in detecting HPT patients. The most critical parameter limiting this study the data is not distributed evenly. In future studies, without manual feature extraction deep learning methods that take an end-to-end approach in detecting HPT patients is planned to investigate their performance.

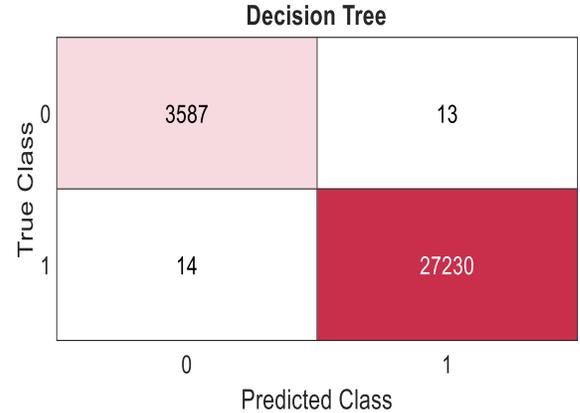

**Figure 2.** Confusion matrice of the highest performance achieved with DT for 5000 length signal

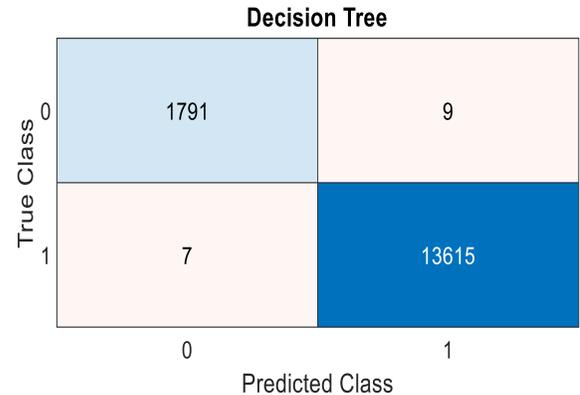

**Figure 3.** Confusion matrice of the highest performance achieved with DT for 10000 length signal

**Table 2:** Performance comparison of segmentations for all EMD based features.

| Decision Tree | Performances (%) | | | | |
|---|---|---|---|---|---|
| | Acc | Rec | Spe | Pre | F1 |
| Five thousand Length | 99.9991 | 99.9994 | 99.6 | 99.9995 | 99.9995 |
| Ten thousand Length | 99.9989 | 99.9994 | 99.5 | 99.9993 | 99.9994 |

**Table 3:** HPT detection engaging ECG signals in the literature. (Acc, accuracy).

| Authors | Methods and Classifiers | Classification | Performance (%) |
|---|---|---|---|
| Khan ve arkadaşları [7] | EMD | k-NN | Acc = 99.4 |
| Rajput ve Ark. [8] | DWT | k-NN, SVM, EBT | Acc=99.95 |
| Soh ve ark. [9] | CNN | FCL | Acc=99.99 |
| Poddar ve ark.[10] | TD and FFT measurements | SVM | Acc=100 |
| **This study(5000 length ECG)** | **EMD basis features** | **DT** | **Acc=99,9991** |
| **This study(10000 length ECG)** | **EMD basis features** | **DT** | **Acc=99,9989** |